\pdfoutput=1
\documentclass{PoS}
\usepackage[utf8]{inputenc}
\usepackage{cite}
\usepackage{graphicx}
\usepackage{amsmath}
\usepackage{amssymb}
\usepackage{slashed}

\allowdisplaybreaks

\def\beq{\begin{equation}}
\def\eeq{\end{equation}}
\def\beqa{\begin{eqnarray}}
\def\eeqa{\end{eqnarray}}

\newcommand{\be}{\begin{equation}}
\newcommand{\ee}{\end{equation}}
\newcommand{\bea}{\begin{eqnarray}}
\newcommand{\eea}{\end{eqnarray}}

\newcommand{\as}{\alpha_s}
\newcommand{\eps}{\epsilon}
\newcommand \slsh [1] {\not\!{#1}}
\newcommand \e {\epsilon}

\title{Universal structure of 
the Drell-Yan process beyond threshold}

\ShortTitle{Drell-Yan beyond threshold}

\author{\speaker{Domenico Bonocore}\thanks{Talk
based on a collaboration with 
  E. Laenen, L. Magnea, S. Melville, L. Vernazza  and  C.D. White. 
}\\
        Nikhef \\
        E-mail: \email{d.bonocore@nikhef.nl}}

\abstract{We review
recent results in the 
 investigation of threshold logarithms at next-to-leading power considering the case of the Drell-Yan cross section at NNLO.
We first show how they can be reproduced with a method of region 
calculation.
Then we move to an approach based on 
soft-collinear factorization, showing that
the entire logarithmic structure can be reproduced
by means of universal functions. 
}

\FullConference{12th International Symposium on Radiative Corrections (Radcor 2015) and LoopFest XIV (Radiative Corrections for the LHC and Future Colliders)\\
		15-19 June, 2015\\
		UCLA Department of Physics \& Astronomy Los Angeles, USA}

\begin{document}

\section{Introduction}

It is well known that cross sections close to the threshold 
are plagued by potentially large logarithms of soft and collinear origin 
that often need to be resummed to all orders in order to 
maintain the predictability of perturbation theory.
For a variety of processes the logarithmic structure
of these cross sections
follows the pattern
 \begin{align}
 \frac{d \sigma}{d \xi} \, = \, \sum_{n = 0}^{\infty}  \alpha_s^n \, 
  \sum_{m = 0}^{2 n - 1} \left[ a_{n m}\,
 \left(\frac{\log^m(\xi)}{\xi}\right)_+\, + \,\, b_{nm} \, \log^m(\xi) 
   \, \,+ \, \, \mathcal O(\xi) \right] \, ,
\label{thresholddef}
\end{align}
where 
 $\xi$ is a dimensionless variable that vanishes in the threshold limit.

The literature about the resummation of the leading terms, which 
are plus-distributions in momentum space, is extensive~\cite{Sterman:1986aj,Catani:1989ne,Contopanagos:1996nh} and relies largely on the \emph{eikonal} approximation.  
 Preliminary studies on the structure of the subleading terms
 have been 
 performed in~\cite{Laenen:2008ux,Laenen:2008gt,Laenen:2010uz,Moch:2009hr,Moch:2009mu,Grunberg:2009yi,Apolinario:2014csa,Almasy:2015dyv}. 
 They typically require a generalization 
 of standard techniques to a subleading degree of approximation 
 (known in the literature as ``\emph{next-to-soft}''
 or ``\emph{next-to-eikonal}''). 
 However, a general
  formalism that resums such contributions is still lacking. 
 In parallel, the subject has been investigated at amplitude level in more formal contexts \cite{White:2011yy,Cachazo:2014fwa,Casali:2014xpa,Bern:2014oka,Luo:2014wea,Larkoski:2014hta,White:2014qia,Bern:2014vva,Broedel:2014fsa,Brandhuber:2015vhm} and with the use of soft-collinear effective theory\cite{Larkoski:2014bxa}.
  Here we report recent results that shed light on this matter \cite{Bonocore:2014wua,Bonocore:2015esa},
 focusing on the specific case of the Drell-Yan process.
  
  This process, at parton level,
  describes the production of an off-shell
  gauge boson of invariant mass $Q^2$ from 
  a quark-antiquark pair of invariant mass $s$.
  Introducing the dimensionless ratio $z=Q^2/s$, 
 the threshold limit is achieved in the limit $z~\to~1$.
 Among all the reasons that make this process
 interesting, probably the most
 relevant one for LHC physics is the similarity with 
  Higgs boson production through gluon-fusion,
   thanks to a common topology of the diagrams 
 involved~\cite{Anastasiou:2015ema, Anastasiou:2014vaa,Li:2014afw}.
   However, there is also a more technical reason for studying it:
    in this process kinematical constraints
  force threshold radiation to be soft 
  and thus the threshold limit corresponds to the vanishing 
  of the energy of the real gluon. However, as we shall see, this
  does not imply that threshold
logarithms are insensitive to the virtual gluon when this is collinear.
 
\begin{figure}[htbp]
\begin{center}
\includegraphics[scale=0.90]{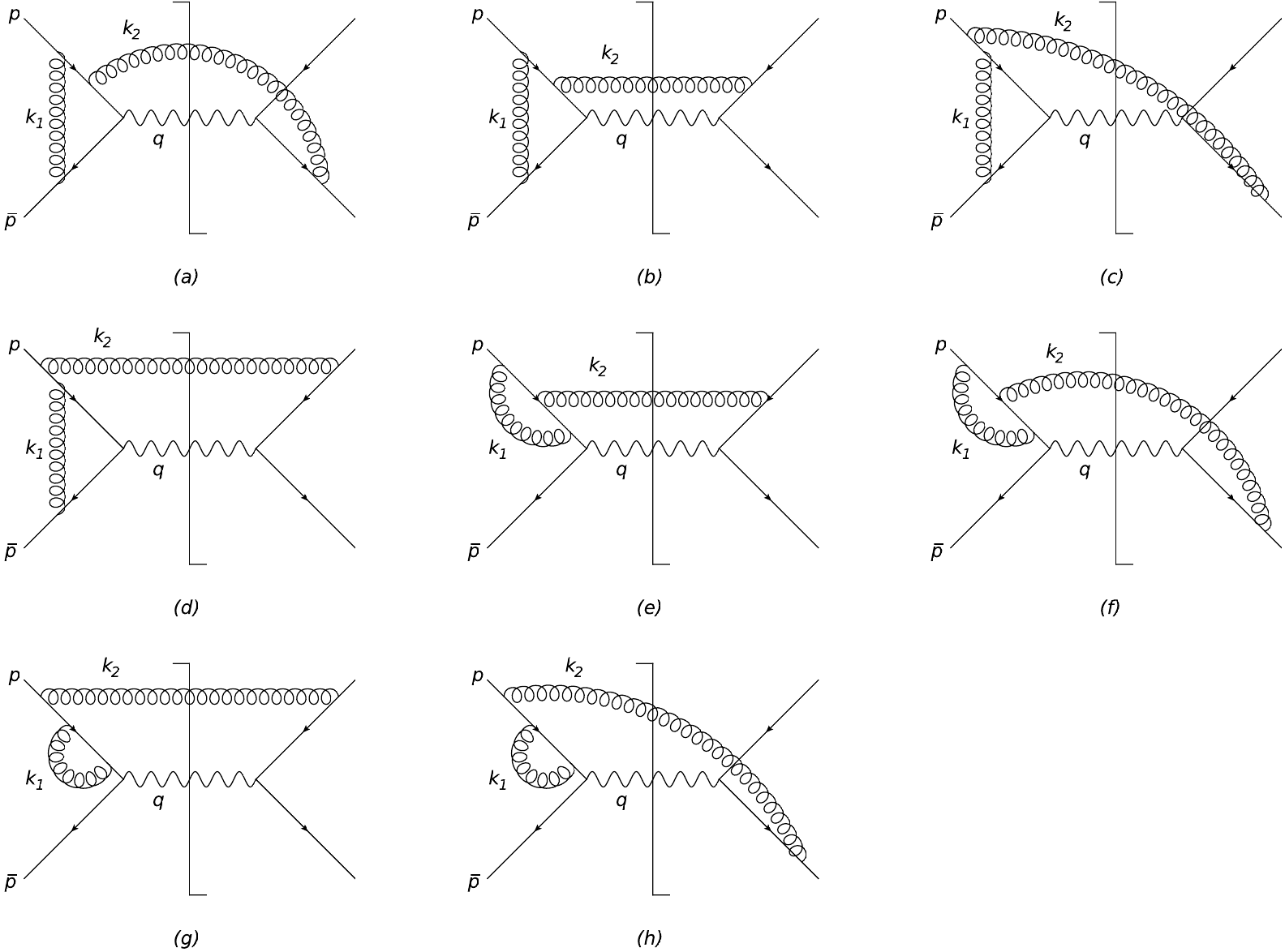}
\caption{Abelian-like diagrams for the real-virtual interference of NNLO Drell-Yan production.
Diagrams obtained by complex conjugation and $p \leftrightarrow \bar p$
are omitted.}
\label{fig:1r1v}
\end{center}
\end{figure}

The case considered here is 
 the $C_F^2$ part of the real-virtual interference at NNLO \cite{Hamberg:1990np}, whose diagrams are represented in Figure~\ref{fig:1r1v}.
While the restriction to the abelian part is merely dictated by
simplicity, in order to minimize technical 
complications, 
the choice of this set of diagrams is related to
the difference between soft (or eikonal) and threshold expansion.
These two do not coincide when there are both real and virtual gluons
since the threshold expansion also includes collinear effects.
Therefore
we cannot use
 factorization techniques based on effective next-to-soft vertices
that turned out to be successful  for diagrams with two real gluons~\cite{Laenen:2010uz}.

To investigate these issues, we consider the contribution of these diagrams to 
the NNLO $K$-factor,
which is defined as 
\begin{align}
 K^{(2)} (z) = \frac{1}{\sigma_B}\frac{d \sigma^{(2)}}{d Q^2} \, ,
 \end{align}
where $\sigma_B$ is the Born cross section and the superscript
stands for NNLO in $\alpha_s$.
In~\cite{Bonocore:2014wua}
the exact result expanded at subleading order in the 
threshold was presented. 
Discarding for simplicity transcendental constants,
 it reads
 \begin{align}
\label{K2tot}
  K^{(2)} (z) & =  \left( \frac{\alpha_s}{4 \pi} C_F\right)^2
  \Bigg[ \frac{32 {\cal D}_0 (z) - 32} {\epsilon^3} + \frac{- 64{\cal
        D}_1 (z) + 48 {\cal D}_0 (z) + 64 L(z) - 96}{\epsilon^2}
    \nonumber \\ 
    &  \hspace{2 cm} + \, \frac{ 64 {\cal D}_2 (z) - 96 {\cal D}_1
      (z) + 128 {\cal D}_0 (z) - 64 L^2 (z) + 208 L(z) -
      196}{ \epsilon} 
      \nonumber \\ 
       & \hspace{2 cm} - \, \frac{128 {\cal
        D}_3 (z)}{3} + 96 {\cal D}_2 (z) - 256 {\cal D}_1 (z) + 256 {\cal
      D}_0 (z) 
      \nonumber \\ 
      & \hspace{2 cm} + \frac{128}{3} L^3 (z) 
      - 232 L^2 (z) + 412
    L(z) -408 \;\;+\, \mathcal O(1-z)\Bigg] \quad , 
\end{align}
where 
 \begin{align}
{\cal D}_m(\xi)=\left(\frac{\log^m(\xi)}{\xi}\right)_+ 
\qquad
\text{and}
\qquad
L(z)=\log(1-z)
\quad .
\end{align}
In 
order to understand the underlying singular (universal)
nature of these logarithms,
in the following sections we review 
how to reproduce their coefficients with two approaches: the method of regions
and a factorization approach.

\section{Analysis with the method of regions}
The method of 
regions~\cite{Beneke:1997zp}
is a powerful tool to 
compute loop integrals in a specific asymptotic limit and 
amounts to expanding the integrand in various regions
 according to its singular behavior.
In the case we want to consider, the 
small parameter $\lambda$ of this
asymptotic limit defines the soft (or eikonal) expansion and
is proportional to the square root of the energy 
of the \emph{real} gluon.
Since we want to perform such an 	expansion 
at subleading order \cite{Bonocore:2014wua}, we distinguish between the leading
eikonal (E) and subleading next-to-eikonal (NE) contributions.
The relevant regions are defined by the 
momentum~$\ell$
of the \emph{virtual} gluon, according to 
the different scaling of its components, which 
in light-cone coordinates reads
\begin{align}
\text{hard} &:\, \ell_h\,\sim\,  (1,1,1)
 \hspace{3.05cm} \text{collinear} :\,\ell_c\,\sim\,  (1,\lambda,\lambda^2)  \notag \\
\text{soft} &: \,\ell_s\,\sim\,  (\lambda^2,\lambda^2,\lambda^2)
\hspace{1.8cm} \text{anticollinear} : \,\ell_{\bar c}\,\sim\, 
(\lambda^2,\lambda,1) \quad.
\end{align}

However, most of the regions are actually zero for some subsets of diagrams. First of all, in the soft region all integrals are scaleless and thus must be set to 
zero\footnote{Setting to zero scaleless integrals in dimensional regularization is one of the prescriptions of the method of regions\cite{Jantzen:2011nz}. Note 
that the soft region is different 
from the traditional definition of the soft function: this one is also represented
by scaleless integrals but, being defined as the VEV of composite operators, needs a further UV renormalization
and thus it is not zero.}.
Then, at E level, (anti)collinear regions are non-zero on a diagram by diagram basis. However, in Feynman gauge\footnote{This choice is consistent with the analysis 
carried out with a factorization approach.}, 
they cancel in every pair of graphs in which the attachments of the two 
gluons are swapped. 
Finally, self-energy diagrams (e)-(h) are made out of
pure collinear regions.
In conclusion, non-vanishing contributions come
from the hard region  $K^{(2)}_{h}$ of 
diagrams (a)-(d) (both at E and NE level)  and from
the collinear region  $K^{(2) }_{c+\bar c}$ of all diagrams (only at NE level). One finds
\begin{align}
  K^{(2) }_{\rm E, \, h} (z) & =  \left( \frac{\as}{4\pi}C_F \right)^2 \,
  \Bigg[ \frac{32 {\cal D}_0(z)}{\eps^3} + \frac{- 64 +48 {\cal D}_0 (z) - 
  64 {\cal D}_1 (z)}{\eps^2} 
  \nonumber \\
  &  \hspace{2 cm}
  + \, \frac{- 96 + 128 {\cal D}_0 (z) - 96 {\cal D}_1 (z) + 64 {\cal D}_2 (z) 
  + 128 L(z)}{\eps} \nonumber \\
  &  \hspace{2 cm}   +\, 256 {\cal D}_0 (z) - 256 {\cal D}_1 (z) + 96 {\cal D}_2 (z) - 
  \frac{128 {\cal D}_3 (z)}{3}
  \nonumber \\
  &  \hspace{2 cm}
   + \, 192 L(z) - 128 L^2 (z)-256 \Bigg]  \label{NNLOHardE} \,\, ,\\
  K^{(2) }_{\rm NE, \, h} (z) & =  \left( \frac{\as}{4\pi} C_F\right)^2 \,
  \bigg[ - \frac{32}{\eps^3} + \frac{16 + 64 L(z)}{\eps^2} + \, \frac{- 80 + 32 L(z) - 64 L^2 (z)}{\eps} 
  \nonumber \\
  &  \hspace{2 cm}
  + \, 160 L(z) - 32 L^2 (z) + \frac{128}{3} L^3 (z) - 128  
  \Bigg]  \label{NNLOHardNE}
\,\, ,\\
 K^{(2), \, {\rm (a)-(d)} }_{\rm NE, \, c + \bar c}(z) \, &= \,
  \left( \frac{\as}{4\pi} C_F\right)^2 \, 
  \Bigg[ - \frac{8}{\eps^2} + \frac{24\, L(z)}{\eps}  - 36\, L^2(z) + 16 \Bigg] 
 \label{NNLOCollinearNEvert} \,\, ,\\
K^{(2), \, {\rm (e)-(h)} }_{\rm NE, \, c + \bar c}(z) \, &= \, 
\left( \frac{\as}{4\pi} C_F\right)^2 \, 
  \Bigg[ - \frac{8}{ \eps^2} + \frac{- 20 +24 L(z)}{  \eps}   + 
  60 L(z) - 36 L^2 (z) - 40 \Bigg] \,\, .
  \label{NNLOCollinearNEext} 
\end{align}
The sum of these four contributions reproduces the full result
of Eq.~(\ref{K2tot}).
Plus-distributions are exclusively captured by the hard function and, as expected, they are entirely reproduced already at
leading order (E) in the soft expansion.
However, the presence of $L(z)$'s in Eq.~(\ref{NNLOHardE}), 
which are subleading in $(1-z)$, shows explicitly the difference between threshold and eikonal (or soft) expansion, and in this case
it is due to corrections to the phase space measure.

 We can investigate further the role of these regions in the context of a na\"ive factorization approach.
 Let us consider the real emission of a next-to-soft 
 gluon of momentum $k$
 from a hard line of momentum $p_i$. Let us assume further that this on-shell gluon factorizes 
 from the hard line 
 and therefore the entire dependence on $k$ can be 
 represented by 
 a universal factor $V$ \cite{Laenen:2008ux, Laenen:2008gt,White:2014qia} that can be expanded at E and NE level
 and reads respectively 
 \begin{align}
\label{VE}
V_E&=- \frac{p_i^{\mu}}{p_i\cdot k} \quad , \\ 
V_{NE}&=\frac{k^{\mu}}{2p_i\cdot k}
-\frac{i\, k_{\nu}\Sigma^{\mu\nu}}{p_i\cdot k} \quad.
\label{VNE}
\end{align}
From these it is clear 
that eikonal interactions, as it is well known, cannot resolve
the spin of the emitting particle (and therefore
they are represented by a scalar vertex). 
NE emissions instead are spin-dependent and contain
the Lorentz generator $\Sigma^{\mu\nu}=\frac{i}{4}\left[\gamma^{\mu},\gamma^{\nu}\right]$.

Working in a na\"ive factorization framework, 
we can use (\ref{VE})-(\ref{VNE}) to compute the 
contribution to the NNLO $K$-factor shown in
Fig.~(\ref{fig:1r1v}). In particular,
 diagrams can be written as a one-loop amplitude
(either a vertex form factor or a self energy correction)
times the real emission at E or NE level.
However, following this procedure, most of the diagrams vanish,
either because after taking the spinor trace we encounter $p_i^2=0$
or
because the self energy diagram itself is zero.
 The only non vanishing ones would be
diagrams (a) and (c), whose sum matches with the
result from an exact computation.
 Hence,
  all plus-distributions and the leading logarithms
  are reproduced by this na\"ive factorizaion approach.
On the other hand we know from the method of region
analysis that the entire hard region is 
given by diagram (a)-(d).
Therefore, recalling that diagram (d) is zero~\cite{Bonocore:2015esa},
 we conclude that the contributions which are left out 
 in this na\"ive factorization approach are
 the entire collinear region and the hard
  region of diagram (b).
  

In conclusion, with the method of regions we have been able to
reproduce the entire logarithmic structure,
attributing each logarithm to a specific singular region.
Moreover,
it has been possible to identify the
kinematical nature of the terms that break a na\"ive factorization approach.
However, if we want to address an all-order interpretation
for the \emph{entire} logarithmic structure, we need 
to revisit our factorization approach.

\section{A new factorization approach}
The starting point is the soft-collinear factorization formula for scattering amplitudes
in covariant gauge \cite{Dixon:2008gr}. For the quark form factor ${\cal A}$
of external momenta $p_1$ and $p_2$
 it reads
\begin{align}
\label{softcolfac}
  {\cal A} \left( \frac{p_1\cdot p_2}{\mu^2}, \alpha_s (\mu^2), \epsilon \right) & =  
  {\cal H} \Big( \{ p_i \}, \{ n_i \}, \alpha_s (\mu^2), \epsilon \Big)
 \, \times\,  \overline{\cal S} \Big( \{ \beta_i \}, \{ n_i \}, \alpha_s (\mu^2), \epsilon \Big) \nonumber \\ 
  &  \qquad \quad \times \, \prod_{i = 1}^2  \, J_i \Big( p_i, n_i, \alpha_s (\mu^2), 
  \epsilon \Big)   \quad ,
\end{align}
where ${\cal H}$, $J$ and $\overline{\cal S}$ are respectively the
 \emph{hard}, \emph{jet} and \emph{reduced soft} functions.
   Their definitions in terms 
of composite operators can be found e.g. in \cite{ Dixon:2008gr, Gardi:2009qi}, where also an explicit one 
loop expression for them is given.
For each external leg $i$, $n_i$ is an auxiliary factorization vector, used to define
the direction of the Wilson lines that appear in the operator definition. In order to make a closer connection with the method of regions and to simplify the calculation, in the following it will be assumed that $n_i^2=0$.

Building upon the work done in \cite{DelDuca:1990gz}, in \cite{Bonocore:2015esa} this formula has been generalized
to the case of an amplitude ${\cal A}^\mu$ radiating
an extra (next-to-)soft gluon
of momentum $k^{\mu}$
\begin{align}
\label{NEfactor}
  {\cal A}^\mu (p_j, k)  & = \sum_{i = 1}^2 \Bigg[ \, q_i \left( \frac{(2 p_i - k)^\mu}{2 p_i 
  \cdot k - k^2} + G^{\nu \mu}_i \, \frac{\partial}{\partial p_i^\nu} \right) {\cal A} (p_i; p_j) \notag \\
  &  \hspace{1mm} + \, {\cal H} (p_j, n_j) \, \overline{\cal S}(\beta_j, n_j) \, G^{\nu \mu}_i
  \left( J_\nu (p_i, k, n_i) - q_i \, \frac{\partial}{\partial p_i^\nu} J(p_i, n_i) \right)
  \prod_{j \neq i} J(p_j, n_j) \Bigg] \,\, ,
\end{align}
where the tensor $G^{\mu\nu}$ is defined in terms of the Minkowski metric $\eta^{\mu\nu}$ as
\begin{align}
G^{\mu\nu}=\eta^{\mu\nu}-\frac{(2p-k)^{\nu}}{2p \cdot k -k^2}k^{\mu} \quad .
\end{align}
This formula 
shows, up to next-to-soft level, how a radiative amplitude ${\cal A}^\mu$ is related 
to the non-radiative one ${\cal A}$.
The first term in the first line of (\ref{NEfactor}) represents the scalar contribution to a factorizable external emission,
already discussed in equations (\ref{VE})-(\ref{VNE}).
At eikonal level, this is the only term which survives and 
simply describes traditional factorization via an external 
eikonal emission. 
 Then 
there are two derivative contributions
(acting respectively on the full form factor and on the jet function) which were present in the original Low-Burnett-Kroll analysis
\cite{Low:1958sn, Burnett:1967km}.
The last non-derivative term  
was first discussed by Del Duca \cite{DelDuca:1990gz}
and requires the definition
of the \emph{radiative jet} function. 
Physically, it represents a soft emission from the jet function and it is defined as
\begin{align}
 J_\mu \left( p, n, k, \alpha_s(\mu^2), \epsilon \right) u(p) \, = \, 
  \int d^d y \,\, {\rm e}^{ - {\rm i} (p - k) \cdot y} \, \left\langle 0 \left| \,
   \, \Phi_{n} (y, \infty) \, \psi (y) \, j_\mu (0) \, \right| p \right\rangle \, ,
\label{Jmudef}
\end{align}
While this formal definition has been known for a long time, its explicit 
expression was computed only in \cite{Bonocore:2015esa}.
For $n_i^2=0$ it
reads respectively at tree level and one loop
\begin{align}
\label{Jnu0}
 J^{\nu(0)} \left(p, n, k \right) &= - \, \frac{p^\nu}{p \cdot k} + \frac{k^\nu}{2 p \cdot k}
 - \frac{{\rm i} \, k_{\alpha} \Sigma^{\alpha \mu}}{2 p \cdot k} \quad ,\\
  J^{\nu (1)} \left(p, n, k \,; \epsilon \right) & =  \left( 2 p \cdot k \right)^{- \epsilon} 
  \Bigg[ \left(\frac{2}{\epsilon} + 4 + 8 \epsilon \right) \left(\frac{n \cdot k}{p \cdot k}
  \frac{p^\nu}{p \cdot n} - \frac{n^\nu}{p \cdot n} \right)
  - (1 + 2 \epsilon) \, \frac{{\rm i} \, k_\alpha \Sigma^{\alpha \nu}}{p \cdot k} \nonumber \\
  &  + \, \left(\frac{1}{\epsilon}-\frac{1}{2}-3\epsilon\right)
  \frac{k^\nu}{p \cdot k} + \left(1 + 3 \epsilon \right) \left(\frac{\gamma^\nu \slsh{n}}{p \cdot n}
  - \frac{p^\nu \slsh{k} \slsh{n}}{p \cdot k \, p \cdot n} \right)
  \Bigg] + \mathcal O(\e^2, k) \,\, .
\label{eq:nullj} 
\end{align}
We observe that the tree level contribution
is equal to the sum of equation (\ref{VE}) and (\ref{VNE}),
and therefore describes a full spin-dependent factorizable emission. The one loop expression instead is proportional 
to $\left( 2 p \cdot k \right)^{- \epsilon}$, which is the typical
collinear scaling. Moreover we note that at one loop the spin dependent
part is subleading in $\epsilon$ and therefore its contribution
will affect only subleading logarithms of the $K$-factor. 

A number of simplifications can be done. Following a renormalization group argument,
it is possible to perform a bare calculation
ignoring UV counterterms. 
This simplifies considerably  
the factorization formula:
without UV counterterms, for $n_i^2=0$, 
the jet function becomes simply
equal to the identity since
its radiative corrections vanish.
Then, in order to make a more direct connection
with the method of region analysis, we can set
$n_i$ to be equal to the anti-collinear direction of the leg $i$
(i.e. $n_1=p_2$ and $n_2=p_1$). This is physically motivated
by the fact that the vector $n_i$ has been introduced via the Wilson line as a replacement for the parton of momentum $p_i$. 	
Finally, we can expand the factorization formula at one loop and make use of (\ref{Jnu0}). Thus we get
\begin{align}
\label{NEfactor5}
  {\cal A}^{\mu, (1)} (p_j, k) & =  \sum_{i = 1}^2 \Bigg[ \left( 
  \frac{p^{\mu}_i}{p_i\cdot k_2}-  \frac{\slsh{k_2}\gamma^{\mu}}{2p_i\cdot k_2}
  \right){\cal A}^{(1)} (p_i; p_j)+   \left(G^{\nu \mu}_i \frac{\partial}{\partial p_i^\nu} 
   \right) {\cal A}^{(1)} (p_i; p_j) \nonumber \\ 
  &  \hspace{2cm} + \,  
  J^{\mu\,(1)} (p_i, k) {\cal A}^{(0)}(p_i; p_j) \Bigg] \quad .
\end{align}
This is the final formula we want to use to reconstruct the NNLO $K$-factor. 
It is made out of three main contributions:
\begin{itemize}
\item a factorized emission external to the one loop form factor,
\item a derivative of the one loop form factor,
\item the one loop radiative jet.
\end{itemize}
In the following we present the contribution to the $K$-factor separately for these three contributions,
comparing them with the method of regions analysis,
and showing that the entire logarithmic structure
of Eq.~(\ref{K2tot}) is reproduced.

\subsection{External contribution}
The set of logarithms coming from the 
form factor dressed by an external emission reads
\begin{align}
\label{K2FF}
  K^{(2)}_{\rm ext} (z) & =  \left( \frac{\alpha_s}{4\pi} C_F \right)^2 \Bigg\{
  \frac{32}{\epsilon^3} \, \Big[ {\cal D}_0 (z) - 1 \Big] +
  \frac{8}{\epsilon^2} \, \Big[ - 8 {\cal D}_1 (z) + 6 {\cal D}_0 (z) + 8 L(z) - 14 \Big] 
  \nonumber \\
  &  \hspace{2cm} + \, \frac{16}{\epsilon} \, \Big[ 4 {\cal D}_2 (z) - 6 {\cal D}_1 (z) + 
  8 {\cal D}_0 (z) - 4 L^2 (z) + 14 L (z) - 14 \Big] \nonumber \\
  &  \hspace{2cm} - \, \frac{128}{3} {\cal D}_3 (z) + 96 {\cal D}_2 (z) - 256 {\cal D}_1 (z) + 
  256 {\cal D}_0 (z) \nonumber \\
  &  \hspace{2cm} + \, \frac{128}{3} L^3 (z) - 224 L^2 (z) + 448 L (z) - 512 \Bigg\} \quad .
\end{align}
As expected all ${\cal D}$'s of the total result come from this contribution 
(i.e. eikonal emissions factorize).
This term collects the entire factorizable part of the $K$-factor, which from the method of region analysis we found to be equal to  the hard regions of diagrams (a) and (c).

\subsection{Derivative contribution}
The contribution from the derivative of the full form factor gives
\begin{align}
  K^{(2)}_{\partial {\cal A}} (z) \, = \, \left(\frac{\alpha_s}{4\pi} \, C_F \right)^2 
  \Bigg\{ \frac{32}{\epsilon^2} + \frac{16}{\epsilon} \, \Big[ - 4 L(z)  + 3 \Big]
  + 64 L^2 (z) - 96 L (z) + 128 \Bigg\} \, ,
\label{Kderiv}
\end{align}
This term is the one related to original analysis of Low, Burnett and Kroll and it matches with the non-factorizable part of the hard region,
i.e. the hard region of diagram (b).
As expected it contains no ${\cal D}$'s, and thus does not spoil 
eikonal factorization.

\subsection{The radiative jet contribution}
The third and last contribution comes the radiative jet function. Just as the derivative contribution,
it breaks 
na\"ive factorization and reads
\begin{align}
  K_{\rm jet}^{(2)} (z) \,& = \, \left( \frac{\alpha_s}{4\pi} \,C_F \right)^2
  \Bigg\{ - \frac{16}{\epsilon^2} + \frac{4}{\epsilon} \, \Big[ 12 L(z) - 5 \Big] 
  - 72 L^2 (z) + 60 L(z) - 24 \Bigg\} \quad .
\label{Kcoll}
\end{align}
This term matches precisely the total collinear region,
given by the sum of equations~(\ref{NNLOCollinearNEvert}) and (\ref{NNLOCollinearNEext}), and  
it is the source of breakdown of next-to-soft theorems 
at loop level, as first found by \cite{DelDuca:1990gz}.

\section{Conclusions}
We considered the NNLO Drell-Yan $K$-factor as case study for the investigation of threshold logarithms at next-to-leading power.
We first showed how this can be reproduced by a method of region 
calculation, isolating and studying the nature of the terms that break 
a na\"ive factorization.  
Then we moved to an approach based on the operator formalism of soft-collinear factorization, showing that
again the entire logarithmic structure can be reproduced.
Crucial to this analysis was the collinear region and the computation of the radiative jet.
Reproducing all threshold logarithms at next-to-leading power via universal functions is a very promising result that should
pave the way for a full resummation formalism.

\bibliography{ref.bib}

\providecommand{\href}[2]{#2}\begingroup\raggedright\begin{thebibliography}{10}

\bibitem{Sterman:1986aj}
G.~F. Sterman, ``{Summation of Large Corrections to Short Distance Hadronic
  Cross-Sections},''
\href{http://dx.doi.org/10.1016/0550-3213(87)90258-6}{{\em Nucl. Phys.}
  {\bfseries B281} (1987) 310}.

\bibitem{Catani:1989ne}
S.~Catani and L.~Trentadue, ``{Resummation of the QCD Perturbative Series for
  Hard Processes},''
\href{http://dx.doi.org/10.1016/0550-3213(89)90273-3}{{\em Nucl. Phys.}
  {\bfseries B327} (1989) 323}.

\bibitem{Contopanagos:1996nh}
H.~Contopanagos, E.~Laenen, and G.~F. Sterman, ``{Sudakov factorization and
  resummation},'' \href{http://dx.doi.org/10.1016/S0550-3213(96)00567-6}{{\em
  Nucl. Phys.} {\bfseries B484} (1997) 303--330},
\href{http://arxiv.org/abs/hep-ph/9604313}{{\ttfamily arXiv:hep-ph/9604313
  [hep-ph]}}.

\bibitem{Laenen:2008ux}
E.~Laenen, L.~Magnea, and G.~Stavenga, ``{On next-to-eikonal corrections to
  threshold resummation for the Drell-Yan and DIS cross sections},''
  \href{http://dx.doi.org/10.1016/j.physletb.2008.09.037}{{\em Phys. Lett.}
  {\bfseries B669} (2008) 173--179},
\href{http://arxiv.org/abs/0807.4412}{{\ttfamily arXiv:0807.4412 [hep-ph]}}.

\bibitem{Laenen:2008gt}
E.~Laenen, G.~Stavenga, and C.~D. White, ``{Path integral approach to eikonal
  and next-to-eikonal exponentiation},''
  \href{http://dx.doi.org/10.1088/1126-6708/2009/03/054}{{\em JHEP} {\bfseries
  0903} (2009) 054},
\href{http://arxiv.org/abs/0811.2067}{{\ttfamily arXiv:0811.2067 [hep-ph]}}.

\bibitem{Laenen:2010uz}
E.~Laenen, L.~Magnea, G.~Stavenga, and C.~D. White, ``{Next-to-eikonal
  corrections to soft gluon radiation: a diagrammatic approach},''
  \href{http://dx.doi.org/10.1007/JHEP01(2011)141}{{\em JHEP} {\bfseries 1101}
  (2011) 141},
\href{http://arxiv.org/abs/1010.1860}{{\ttfamily arXiv:1010.1860 [hep-ph]}}.

\bibitem{Moch:2009hr}
S.~Moch and A.~Vogt, ``{On non-singlet physical evolution kernels and large-x
  coefficient functions in perturbative QCD},''
  \href{http://dx.doi.org/10.1088/1126-6708/2009/11/099}{{\em JHEP} {\bfseries
  11} (2009) 099},
\href{http://arxiv.org/abs/0909.2124}{{\ttfamily arXiv:0909.2124 [hep-ph]}}.

\bibitem{Moch:2009mu}
S.~Moch and A.~Vogt, ``{Threshold Resummation of the Structure Function
  F(L)},'' \href{http://dx.doi.org/10.1088/1126-6708/2009/04/081}{{\em JHEP}
  {\bfseries 04} (2009) 081},
\href{http://arxiv.org/abs/0902.2342}{{\ttfamily arXiv:0902.2342 [hep-ph]}}.

\bibitem{Grunberg:2009yi}
G.~Grunberg and V.~Ravindran, ``{On threshold resummation beyond leading 1-x
  order},'' \href{http://dx.doi.org/10.1088/1126-6708/2009/10/055}{{\em JHEP}
  {\bfseries 10} (2009) 055},
\href{http://arxiv.org/abs/0902.2702}{{\ttfamily arXiv:0902.2702 [hep-ph]}}.

\bibitem{Apolinario:2014csa}
L.~Apolinário, N.~Armesto, J.~G. Milhano, and C.~A. Salgado, ``{Medium-induced
  gluon radiation and colour decoherence beyond the soft approximation},''
  \href{http://dx.doi.org/10.1007/JHEP02(2015)119}{{\em JHEP} {\bfseries 02}
  (2015) 119},
\href{http://arxiv.org/abs/1407.0599}{{\ttfamily arXiv:1407.0599 [hep-ph]}}.

\bibitem{Almasy:2015dyv}
A.~A. Almasy, N.~A. Lo~Presti, and A.~Vogt, ``{Generalized threshold
  resummation in inclusive DIS and semi-inclusive electron-positron
  annihilation},''
\href{http://arxiv.org/abs/1511.08612}{{\ttfamily arXiv:1511.08612 [hep-ph]}}.

\bibitem{White:2011yy}
C.~D. White, ``{Factorization Properties of Soft Graviton Amplitudes},''
  \href{http://dx.doi.org/10.1007/JHEP05(2011)060}{{\em JHEP} {\bfseries 05}
  (2011) 060},
\href{http://arxiv.org/abs/1103.2981}{{\ttfamily arXiv:1103.2981 [hep-th]}}.

\bibitem{Cachazo:2014fwa}
F.~Cachazo and A.~Strominger, ``{Evidence for a New Soft Graviton Theorem},''
\href{http://arxiv.org/abs/1404.4091}{{\ttfamily arXiv:1404.4091 [hep-th]}}.

\bibitem{Casali:2014xpa}
E.~Casali, ``{Soft sub-leading divergences in Yang-Mills amplitudes},''
  \href{http://dx.doi.org/10.1007/JHEP08(2014)077}{{\em JHEP} {\bfseries 08}
  (2014) 077},
\href{http://arxiv.org/abs/1404.5551}{{\ttfamily arXiv:1404.5551 [hep-th]}}.

\bibitem{Bern:2014oka}
Z.~Bern, S.~Davies, and J.~Nohle, ``{On Loop Corrections to Subleading Soft
  Behavior of Gluons and Gravitons},''
  \href{http://dx.doi.org/10.1103/PhysRevD.90.085015}{{\em Phys. Rev.}
  {\bfseries D90} no.~8, (2014) 085015},
\href{http://arxiv.org/abs/1405.1015}{{\ttfamily arXiv:1405.1015 [hep-th]}}.

\bibitem{Luo:2014wea}
H.~Luo, P.~Mastrolia, and W.~J. Torres~Bobadilla, ``{Subleading soft behavior
  of QCD amplitudes},''
  \href{http://dx.doi.org/10.1103/PhysRevD.91.065018}{{\em Phys. Rev.}
  {\bfseries D91} no.~6, (2015) 065018},
\href{http://arxiv.org/abs/1411.1669}{{\ttfamily arXiv:1411.1669 [hep-th]}}.

\bibitem{Larkoski:2014hta}
A.~J. Larkoski, ``{Conformal Invariance of the Subleading Soft Theorem in Gauge
  Theory},'' \href{http://dx.doi.org/10.1103/PhysRevD.90.087701}{{\em Phys.
  Rev.} {\bfseries D90} no.~8, (2014) 087701},
\href{http://arxiv.org/abs/1405.2346}{{\ttfamily arXiv:1405.2346 [hep-th]}}.

\bibitem{White:2014qia}
C.~D. White, ``{Diagrammatic insights into next-to-soft corrections},''
  \href{http://dx.doi.org/10.1016/j.physletb.2014.08.041}{{\em Phys. Lett.}
  {\bfseries B737} (2014) 216--222},
\href{http://arxiv.org/abs/1406.7184}{{\ttfamily arXiv:1406.7184 [hep-th]}}.

\bibitem{Bern:2014vva}
Z.~Bern, S.~Davies, P.~Di~Vecchia, and J.~Nohle, ``{Low-Energy Behavior of
  Gluons and Gravitons from Gauge Invariance},''
  \href{http://dx.doi.org/10.1103/PhysRevD.90.084035}{{\em Phys. Rev.}
  {\bfseries D90} no.~8, (2014) 084035},
\href{http://arxiv.org/abs/1406.6987}{{\ttfamily arXiv:1406.6987 [hep-th]}}.

\bibitem{Broedel:2014fsa}
J.~Broedel, M.~de~Leeuw, J.~Plefka, and M.~Rosso, ``{Constraining subleading
  soft gluon and graviton theorems},''
  \href{http://dx.doi.org/10.1103/PhysRevD.90.065024}{{\em Phys. Rev.}
  {\bfseries D90} no.~6, (2014) 065024},
\href{http://arxiv.org/abs/1406.6574}{{\ttfamily arXiv:1406.6574 [hep-th]}}.

\bibitem{Brandhuber:2015vhm}
A.~Brandhuber, E.~Hughes, B.~Spence, and G.~Travaglini, ``{One-Loop Soft
  Theorems via Dual Superconformal Symmetry},''
\href{http://arxiv.org/abs/1511.06716}{{\ttfamily arXiv:1511.06716 [hep-th]}}.

\bibitem{Larkoski:2014bxa}
A.~J. Larkoski, D.~Neill, and I.~W. Stewart, ``{Soft Theorems from Effective
  Field Theory},'' \href{http://dx.doi.org/10.1007/JHEP06(2015)077}{{\em JHEP}
  {\bfseries 06} (2015) 077},
\href{http://arxiv.org/abs/1412.3108}{{\ttfamily arXiv:1412.3108 [hep-th]}}.

\bibitem{Bonocore:2014wua}
D.~Bonocore, E.~Laenen, L.~Magnea, L.~Vernazza, and C.~D. White, ``{The method
  of regions and next-to-soft corrections in Drell-Yan production},''
  \href{http://dx.doi.org/10.1016/j.physletb.2015.02.008}{{\em Phys.Lett.}
  {\bfseries B742} (2015) 375--382},
\href{http://arxiv.org/abs/1410.6406}{{\ttfamily arXiv:1410.6406 [hep-ph]}}.

\bibitem{Bonocore:2015esa}
D.~Bonocore, E.~Laenen, L.~Magnea, S.~Melville, L.~Vernazza, {\em et~al.}, ``{A
  factorization approach to next-to-leading-power threshold logarithms},''
  \href{http://dx.doi.org/10.1007/JHEP06(2015)008}{{\em JHEP} {\bfseries 1506}
  (2015) 008},
\href{http://arxiv.org/abs/1503.05156}{{\ttfamily arXiv:1503.05156 [hep-ph]}}.

\bibitem{Anastasiou:2015ema}
C.~Anastasiou, C.~Duhr, F.~Dulat, F.~Herzog, and B.~Mistlberger, ``{Higgs Boson
  Gluon-Fusion Production in QCD at Three Loops},''
  \href{http://dx.doi.org/10.1103/PhysRevLett.114.212001}{{\em Phys. Rev.
  Lett.} {\bfseries 114} (2015) 212001},
\href{http://arxiv.org/abs/1503.06056}{{\ttfamily arXiv:1503.06056 [hep-ph]}}.

\bibitem{Anastasiou:2014vaa}
C.~Anastasiou, C.~Duhr, F.~Dulat, E.~Furlan, T.~Gehrmann, F.~Herzog, and
  B.~Mistlberger, ``{Higgs boson gluon–fusion production at threshold in
  N$^3$LO QCD},'' \href{http://dx.doi.org/10.1016/j.physletb.2014.08.067}{{\em
  Phys. Lett.} {\bfseries B737} (2014) 325--328},
\href{http://arxiv.org/abs/1403.4616}{{\ttfamily arXiv:1403.4616 [hep-ph]}}.

\bibitem{Li:2014afw}
Y.~Li, A.~von Manteuffel, R.~M. Schabinger, and H.~X. Zhu, ``{Soft-virtual
  corrections to Higgs production at N$^3$LO},''
  \href{http://dx.doi.org/10.1103/PhysRevD.91.036008}{{\em Phys. Rev.}
  {\bfseries D91} (2015) 036008},
\href{http://arxiv.org/abs/1412.2771}{{\ttfamily arXiv:1412.2771 [hep-ph]}}.

\bibitem{Hamberg:1990np}
R.~Hamberg, W.~L. van Neerven, and T.~Matsuura, ``{A Complete calculation of
  the order $\alpha-s^{2}$ correction to the Drell-Yan $K$ factor},''
  \href{http://dx.doi.org/10.1016/0550-3213(91)90064-5}{{\em Nucl. Phys.}
  {\bfseries B359} (1991) 343--405}.
[Erratum: Nucl. Phys.B644,403(2002)].

\bibitem{Beneke:1997zp}
M.~Beneke and V.~A. Smirnov, ``{Asymptotic expansion of Feynman integrals near
  threshold},'' \href{http://dx.doi.org/10.1016/S0550-3213(98)00138-2}{{\em
  Nucl.Phys.} {\bfseries B522} (1998) 321--344},
\href{http://arxiv.org/abs/hep-ph/9711391}{{\ttfamily arXiv:hep-ph/9711391
  [hep-ph]}}.

\bibitem{Jantzen:2011nz}
B.~Jantzen, ``{Foundation and generalization of the expansion by regions},''
  \href{http://dx.doi.org/10.1007/JHEP12(2011)076}{{\em JHEP} {\bfseries 12}
  (2011) 076},
\href{http://arxiv.org/abs/1111.2589}{{\ttfamily arXiv:1111.2589 [hep-ph]}}.

\bibitem{Dixon:2008gr}
L.~J. Dixon, L.~Magnea, and G.~F. Sterman, ``{Universal structure of subleading
  infrared poles in gauge theory amplitudes},''
  \href{http://dx.doi.org/10.1088/1126-6708/2008/08/022}{{\em JHEP} {\bfseries
  0808} (2008) 022},
\href{http://arxiv.org/abs/0805.3515}{{\ttfamily arXiv:0805.3515 [hep-ph]}}.

\bibitem{Gardi:2009qi}
E.~Gardi and L.~Magnea, ``{Factorization constraints for soft anomalous
  dimensions in QCD scattering amplitudes},''
  \href{http://dx.doi.org/10.1088/1126-6708/2009/03/079}{{\em JHEP} {\bfseries
  03} (2009) 079},
\href{http://arxiv.org/abs/0901.1091}{{\ttfamily arXiv:0901.1091 [hep-ph]}}.

\bibitem{DelDuca:1990gz}
V.~Del~Duca, ``High-energy bremsstrahlung theorems for soft photons,''
\href{http://dx.doi.org/10.1016/0550-3213(90)90392-Q}{{\em Nucl. Phys.}
  {\bfseries B345} (1990) 369--388}.

\bibitem{Low:1958sn}
F.~E. Low, ``{Bremsstrahlung of very low-energy quanta in elementary particle
  collisions},''
\href{http://dx.doi.org/10.1103/PhysRev.110.974}{{\em Phys. Rev.} {\bfseries
  110} (1958) 974--977}.

\bibitem{Burnett:1967km}
T.~H. Burnett and N.~M. Kroll, ``{Extension of the low soft photon theorem},''
\href{http://dx.doi.org/10.1103/PhysRevLett.20.86}{{\em Phys. Rev. Lett.}
  {\bfseries 20} (1968) 86}.

\end{thebibliography}\endgroup

\end{document}